# WS$_2$-QDs Decorated RGO Lattice on e-textile: Development of Ultrasensitive Wearable Quantum Thermometer

**Abid[1], Poonam Sehrawat[1], C. M. Julien[2], S.S. Islam[1] (*)**

[1]Centre for Nanoscience and Nanotechnology,
Jamia Millia Islamia (A Central University), New Delhi-110025, India.

[2]Institut de Minéralogie, de Physique des Matériaux et de Cosmochimie (IMPMC), Sorbonne Université, CNRS-UMR 7590, 4 place Jussieu, 75252 Paris, France

(*) Corresponding Author Email: sislam@jmi.ac.in; Tel.: + +91 (11) 26987153.

**Abstract:**

We report the fabrication and human trial of a novel wearable temperature sensor based on $WS_2$-QDs/RGO; which performs instant measurement like thermometer in a wide temperature range: 77K-398 K, in both static- and instant mode. The device is simple, scalable, flexible and cost-effective, where nanoscience and technology played a vital role behind its concept and realization. The $WS_2$-QDs/RGO heterostructure is developed by decorating $WS_2$-QDs on pre-RGO coated cotton textile. In static mode, the crucial parameters such as temperature coefficient of resistance (TCR) and thermal hysteresis ($H_{th}$) were analyzed in depth to get the intricate mechanism behind the working of a temperature sensor; and check its worthiness to be a better candidate in the field of temperature sensor. Temperature sensing data at both high- and low temperatures are very much encouraging; and endorses its viability.

In case of instant measurement mode, the sensor works like a thermometer in both high- and low temperature ranges for its application in incubator, and body temperature monitoring. TCR and $H_{th}$ were found nearly unchanged, and independent of the measurement mode; but there is a huge change in case of response- and recovery time which are of the order of few seconds. Human trial is conducted to make reliable and hassle free temperature monitoring like thermometer where the sensor device is found capable to measure accurate body temperature with exceptional resolution i.e. the minimum change in temperature the device can measure is ~0.06K in addition to, fast response- and recovery time ~1.4 s and 1.7 s respectively. In every sense, the developed sensor has exhibited highest degree of superiority vis-à-vis its counterpart commercial thermometer used in healthcare**.** Besides, the device has passed through all deformation test successfully and proved its mettle. This sensor device proved its flexibility and stability under various mechanical deformation(s), showing its promising potential for future generation wearable health monitoring devices. To the best of our knowledge, this is the first report on $WS_2$ in general, and $WS_2$-QDs, in specific, based temperature sensing device and its operational demonstration as of now.

**Introduction**

Highly accurate temperature measurement is an important concern not only for health monitoring applications, but also in our present-day industries be it automobiles, consumer electronics, pharmaceutics, etc. [1-3]. The performance of a temperature sensor is primarily determined from some physical parameters, out of which temperature coefficient of resistance (TCR) [4-8] as well as thermal hysteresis loss ($H_{th}$) [9-10] are crucial and further decide the remaining sensor characteristics including response time, recovery time, operating temperature range, resolution, stability, etc. Other equally important desirable features of a temperature sensor include low power consumption, flexibility, and reduced dimensionality [11,12]. The astounding electronic characteristics of layered two dimensional (2D) materials have shown the viability of their use in nanoscale to get sensor performance better than other devices of similar size [13]. Flexible electronic sensors are gaining increasing popularity for their use in several devices such as display [14], biomedicine [13,15,16], robots, [17] as well as environment monitoring [18,19] devices. Using flexible substrates considerably reduces the device, and allow it to conform, bend or roll into desired shapes, thereby rendering economic mass scale fabrication possible.

The increasing popularity of multifunctional wearable e-textiles stems from their ability to make radical lifestyle changes, such as, safety and health aspects coupled with comfort. As people are becoming more and more aware regarding health and fitness, body-wearable temperature sensors are gaining huge audience. Real-time temperature measurement has been shown to indicate various ailments including heart attack. In addition to this, temperature monitoring is frequently undertaken wherever physical activities directly link to the body temperature. Further, it is often neither possible nor desirable to use conventional temperature sensing devices such as regular thermometers, e.g., it appears quite appealing to wear a t-shirt that might measure your body temperature besides performing its conventional aesthetic function. In this regard, various bendable substrates, e.g., polydimethylsiloxane (PDMS) [20], polyethylene terephthalate (PET) [21], papers [22], textile or polyimide (PI) [23], etc., have been employed in flexible device fabrication. With increasing acceptance of smart living, comes the increasing use of temperature sensors as well. Moreover, the present industrial revolution, the 'Internet of Things or IOT', is further poised to exponentially grow the market of flexible wearable sensors [21].

Yet, there are a lot of obstacles before e-textile technology: it has to replace the existing silicon wafer-based technology and build its own commercial market with standard production techniques, offering reproducibility and reliability of results. As of now, e-textile is still limited to laboratory scale production and seems to be few more years away before we include it in our day to day clothing [24].

Resistive temperature devices (RTDs) belong to extensively employed temperature sensor types because of fast response, great accuracy and acceptable stability [25]. The other type of sensors being used are thermocouples [26], infrared detectors [27,28] and mercury-based thermometers [29]. Most of the reports available on flexible temperature sensor development involve the use of various conducting/sensing elements on flexible substrates, such as thin films of heat sensitive metals [30], graphene films with variable number of individual layers [31], and carbon nanotubes (CNTs) [32]. Recently, graphene nanowalls and polyaniline fibers, offering relatively high TCR values, have been studied as potential temperature sensors. Nevertheless, poor linearity limits their development in resistance temperature detectors (RTDs) [33]. As a result, most of the materials reported for RTDs include metals, e.g., Pt, being chemically inert with moderate resistivity ($1.6 \times 10^{-8}$ Ω m), linearity and sufficient TCR value (+0.39 %/°C at 20 °C) [34]. However, its low TCR results in low sensitivity and resolution as well, and also poor sensing performance. Besides, it is expensive and scarce, which necessitates the development of alternate materials as temperature sensors.

As two-dimensional structures, semiconducting transition-metal dichalcogenides (TMDCs), especially $MoS_2$, $WS_2$, and $WSe_2$, are desirable materials for use in electronic devices because of their relatively higher charge carrier mobilities and sufficient bandgaps [35]. Monolayer $WS_2$ demonstrates direct bandgap of ~2.0 eV [35]. Whereas theoretical simulations suggest highest carrier mobility in $WS_2$ because of the reduced carrier effective mass [35]. It is rather easily exfoliated into mono- to few layers and lattice vibrational studies predict in-plane heat transportation akin to graphene [36]. Unfortunately, thermal conductivity of $WS_2$ bulk is ~140 W/mK [37], which is not substantial even in few layered flakes as observed in $MoS_2$ [38]; thereby limiting the development of a temperature sensor solely on its own.

A heterostructure is expected to overcome these challenges based on the premise that $WS_2$ can diffuse extra free carriers in a transport/tunneling medium which might offer good thermal conductivity for rapid thermal transport, thus, improving sensing property. Hybrid structure of $WS_2$-quantum dots ($WS_2$-QDs) and reduced graphene oxide (RGO) is conceived in the present work, where $WS_2$-QDs will release excess free charge carriers (because of its low thermal activation energy ~meV) in the moderate to high conducting (both electrical and thermal) medium for its transportation to be collected as carriers at the terminal electrodes. RGO, another carbon derivative, is desirable due to easy processing unlike graphene synthesis, economic, high yield and on top of it, offers graphene like electrical, electronic and thermal characteristics [39]. Therefore, a transport medium that possesses such characteristics obviously is a good candidate to be associated with a suitable carrier source to make an efficient temperature sensor.

In this work, we have developed as well as demonstrated through human trial, the attributes of an ultra-sensitive wearable temperature sensor which operates in a wide temperature range 77-398K. A bare RGO temperature sensor has been studied in parallel as a reference to assess the sole contribution of $WS_2$-QDs upon decorated on RGO lattice. Both the bare RGO- and $WS_2$-QDs/ RGO based temperature sensors exhibited a negative- and positive temperature coefficient of resistance (TCR) at high- and low temperature ranges respectively, and it endorses their semiconducting nature. Temperature measurements were conducted in two modes- static and instant. The static mode was employed to get the first hand idea on the working of a temperature sensor from the point of physics and chemistry of the heterostructure used; whereas the second one to check the effectiveness of the sensor while in use. The sensing data of both the modes are summarized in Table II and III; while highlighting the superior aspects of $WS_2$-QD/RGO sensor. When the sensor ($WS_2$-QD/RGO) was employed to measure instantaneous temperature just like thermometer (in both the temperature ranges), the sensor parameters were enormously improved that lead to fastest data recording of the device and proved worth for commercial applications such as incubators and solid-state thermometers. Human trial was conducted to monitor human body temperature like thermometer, and it was found that the sensor device is capable to measure a minimum temperature of 0.06 K; showing its efficacy in terms of sensitivity as well as resolution. This developed sensor has shown its flexibility and stable behaviour while mechanical deformation(s) tests such as bending, twisting, stretching etc. were conducted. The attributes of flexibility, high TCR (sensitivity), low thermal hysteresis, and excellent resolution both in static and instant modes as well, in addition to cost-effectiveness of the novel temperature sensor (quantum thermometer); indicate promising potential for future wearable health monitoring devices.

## 1. EXPERIMENTAL SECTION

### 1.1 Materials

Analytical grade graphite powder (99%, 300 mesh, Alfa Aesar), $NaNO_3$ (CDH Fine Chemicals, India), bulk $WS_2$ powders, $KMnO_4$, $H_2SO_4$ (98% conc.), $H_2O_2$, HCl, N-Methyl-2-pyrrolidone (NMP) (Sigma Aldrich) are procured commercially to use for synthesizing GO and $WS_2$-QDs without any pretreatment. Deionized (DI, Millipore system, ~18.2 MΩ) water was used for all experimental purposes.

### 1.2 Synthesis of reduced graphene oxide (RGO) and tungsten disulphide-quantum dots ($WS_2$-QDs)

Graphite oxide (GO) was obtained via modified Hummers process taking natural graphite flakes as precursor as described elsewhere [40]. The synthesized GO (150 mg) was suspended in DI water (20 ml) in bath sonicator up to 50 minutes below 30° C, to obtain a homogenous solution. The graphene oxide coating on cotton was done by dip and drying technique, wherein the cotton cloth was first dipped in graphene oxide solution, and subsequently dried in a hot air electric oven operated at 60 °C for 2 h. Following this, both sides of dried GO-coated cotton fabric were ironed for 20 min using a conventional electric iron set at ~150 °C, such that GO was thermally reduced to RGO. The temperature of the iron or the reduction process was set to a value high enough to reduce the GO and low enough to avoid degradation of the cotton fabric substrate. The whole process is schematically illustrated in Figure. 1. Thus, obtained RGO coated cotton fabric is now available for use in both temperature sensing as well as heterostructure development with $WS_2$-QDs.

The synthesis of $WS_2$-QDs has been reported elsewhere [41]. In brief, $WS_2$ powders (2 g) was mixed in Dimethylformamide (DMF) (150 mL) by sonicating at 30 °C for 3 h. This solution was further processed for 10 min in a probe sonicator for more exfoliation.

Afterwards, top 2/3 solution was decanted into round bottom flask and vigorous stirring was maintained for 16 h at 140 °C; followed by centrifugation of the suspension at 5000 rpm for 5 min, and the harvested yellow supernatant incorporated $WS_2$-QDs. The $WS_2$-QDs (1mL) was dropwise coated on a 1 × 1 cm² piece of RGO cotton cloth through a micropipette without replacing the sensor from the sample holder chuck.

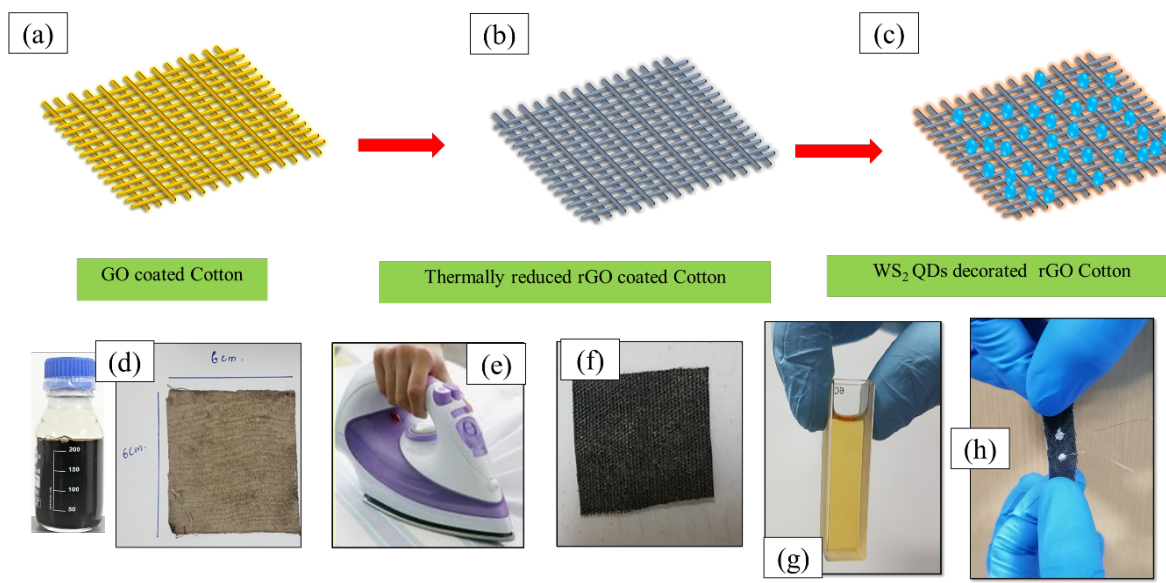

Figure 1. (a-c) Schemes illustrating the development of $WS_2$-QDs decorated RGO-cotton based temperature sensor. Digital photographs of (d) GO coated cotton cloth, (e) Iron employed to thermally reducing the GO, (f) thermally reduced GO (RGO) coated cotton cloth, (g) $WS_2$-QDs, and (h) $WS_2$-QDs/RGO coated developed sensor.

## 2. Characterization*

Figure 2 (a-c) incorporates field emission scanning electron microscopy (FESEM, Nova Nano SEM 450, FEI)) micrographs of natural cotton textile in which cotton threads weaved like a mesh of twisted fibers with smooth and clean surface. The characterization studies were described in detail in Ref. 41. Figure 2 (d-f) shows the GO coated cotton, where GO flakes wrapped around cotton threads are seen. Wrinkles and rough morphology over the surface (Figure 2(g)) clearly indicate the presence of GO on the top of the cotton. Figure 2 (h-j) shows the image of the sample after dropping $WS_2$-QDs solution on to the cotton surface. Figure 2 (k) is the zoomed SEM image of $WS_2$-QDs/RGO cotton, roughly indicating the size of the QDs to be ~3-7 nm. Figure 3(a) displays the TEM (JEOL JEM F-200) image of the wrinkled GO flake. Figure 3(b) is the TEM image of bulk $WS_2$ flake sized approximately 2 µm, where multilayer stacked $WS_2$ flakes look darker against the background. Solvothermal treatment transforms bulk $WS_2$ (2D) to $WS_2$-QDs (0D) of 3-7 nm dimension, and the QDs are shown by yellow arrow in Figure 3(c). The crystalline nature of the QDs are visible in HRTEM image of Figure 3(d); zoomed further to distinctly show the crystalline planes (Figure 3(e)).

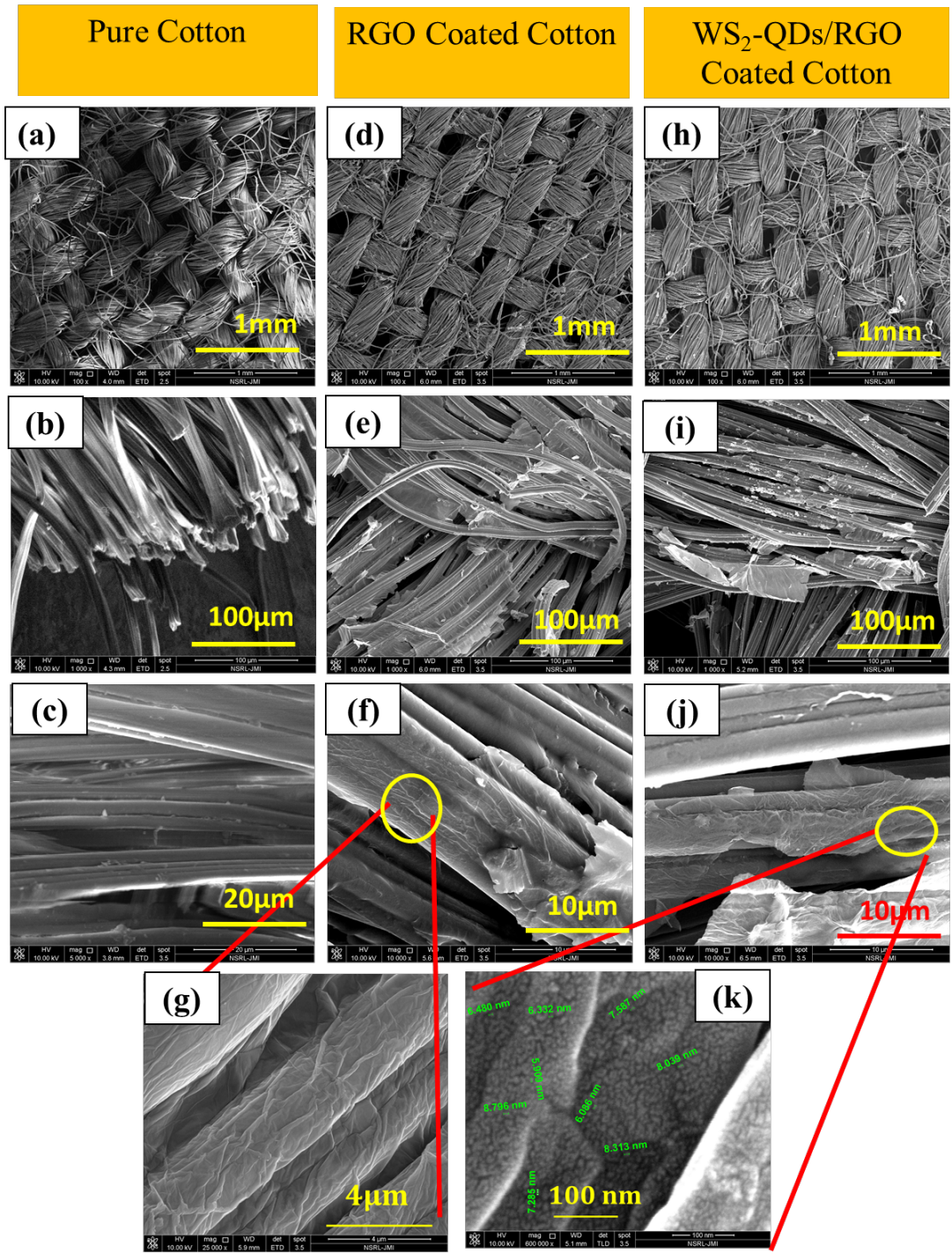

Figure 2. (a-c) Cleaned cotton cloth at different magnifications, (d-f) GO coated cotton, (g) wrapping of GO on the cotton, (h-j) WS$_2$-QDs covered RGO cotton, and (k) zoomed image of the WS$_2$-QDs covered RGO cotton [41].

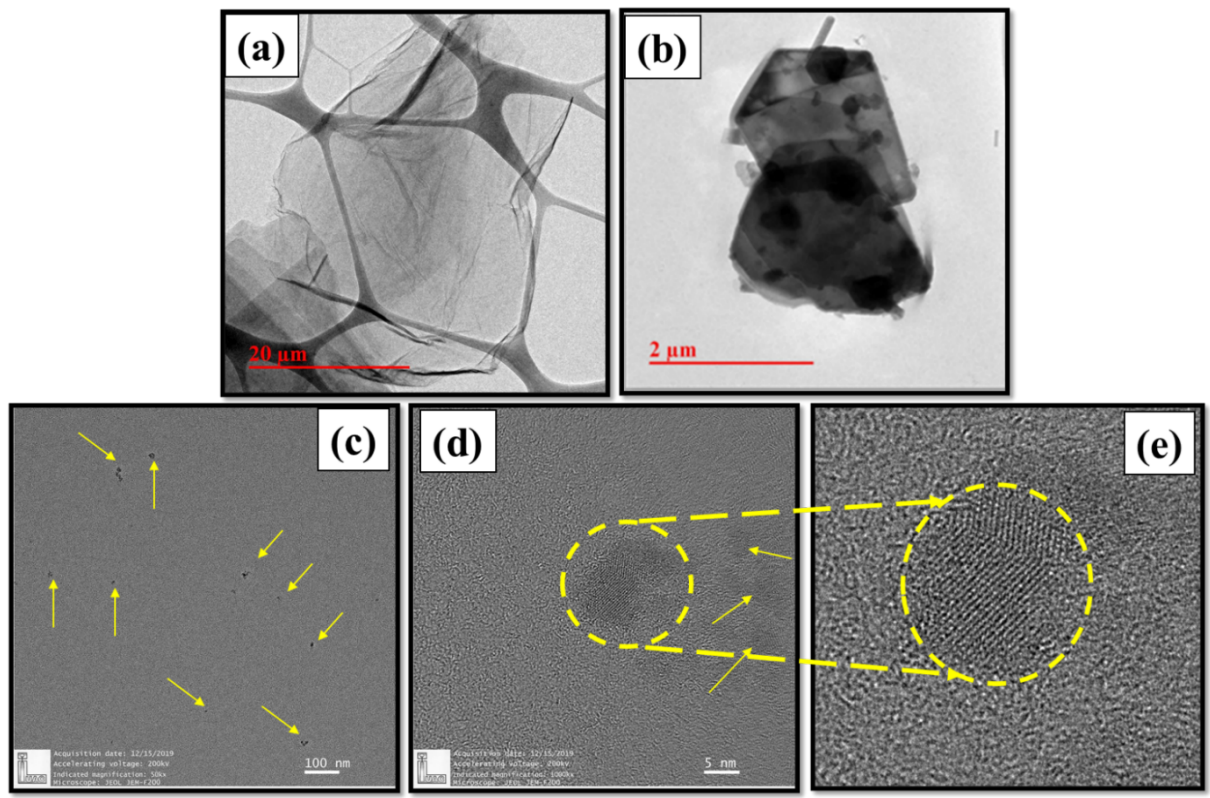

Figure 3. TEM imaging of (a) GO flake, (b) bulk WS$_2$, and (c-e) the WS$_2$-QDs at various magnification levels [41].

The UV-Vis (Agilent carry 100) spectroscopy results of RGO and WS$_2$-QDs are produced in in Figure 4 and detailed in Ref. [41]. For GO sample, peak spotted at 234 nm relates to π-π* transition of the $sp^2$- C-atoms. In RGO spectra, the peak at 268 nm indicates n-π* transitions of the $sp^2$- carbons, which is slightly red shifted from its usual peak at 234 nm; attributable to the restoration of C=C bonds in RGO. The absorption spectra of WS$_2$-QDs were also investigated (Figure 4). A strong absorption band is observed in the UV range (blue curve) which is associated with enhanced quantum confinement effects in smaller nanoparticles in comparison to bulk WS$_2$ nanosheets. Figure 4 (b) shows the schematic diagram of usual electronic transitions in different materials.

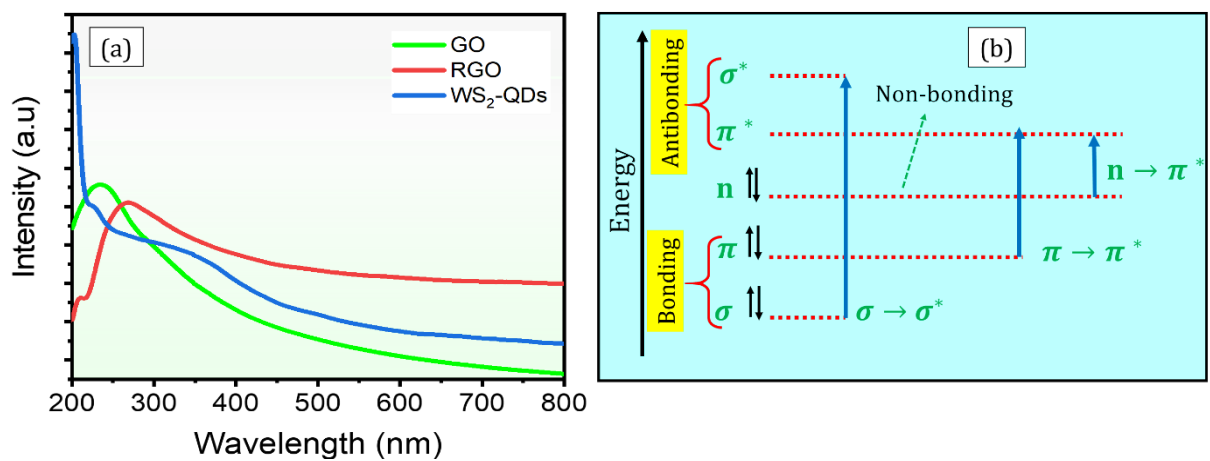

Figure 4. (a) UV-Vis results of GO, RGO, and WS$_2$-QDs, [41] and (b) Molecular transitions in different bonding- anti-bonding states.

The crystal structure of WS$_2$-QDs was probed via X-ray powder diffraction (XRPD). As shown in Figure 5, bulk WS$_2$ powders give strong diffraction peak at 2θ =14.5°, associated with (002) plane and reflections with lower intensity at 29.1, 32.9, 33.7, 39.7, 44.2, 49.9 and 59.9°, corresponding to the (004), (100), (101), (103), (006), (105), and (008) planes, respectively. In sonicated samples, most of these peaks disappeared therefore, endorsing the exfoliation of bulk WS$_2$ to a great degree, as complimented by HRTEM images (Figure 3). WS$_2$-QDs shows relatively weak diffraction patterns, with most of the XRD peaks disappearing due to enhanced exfoliation. Theoretically, *d*-spacing of crystals determines the peak position, and if it is assumed that entire WS$_2$ bulk is turned into monolayer and have no inter-layer interaction, then there would be no peak or signal on XRD spectra. Rao et al. [42] has demonstrated that the characteristic peak disappears when WS$_2$ powders are exfoliated into monolayer like nanosheets. The absence of peaks in WS$_2$-QDs imply the monolayer nature of the exfoliated nanosheets and the small peaks perhaps appear because of the impurity diffractions. The XRD pattern of RGO (Figure 5) displays a broad peak located at

~24.6°, that indicates the effective conversion of GO to reduced graphene oxide [43]. The XRD studies were also discussed in detail elsewhere [41].

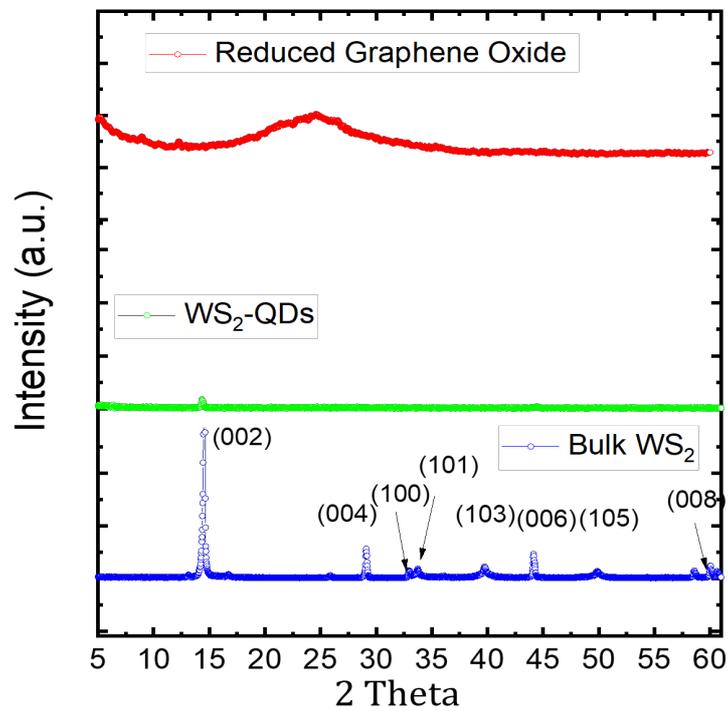

Figure 5. X-ray powder diffraction (XRPD) patterns of RGO, bulk WS$_2$ and WS$_2$-QDs.

Raman spectroscopy is a versatile non-destructive tool popularly known for the investigatation of 2D-materials. Raman spectrophotometer (LabRAM HR800 HORIBA JY) was used to investigate RGO, WS$_2$-QDs, and WS$_2$-QD/RGO independently. Raman spectrum of RGO (Figure 6) shows its characteristic D-band at 1352.6 cm$^{-1}$ and G-band at 1594 cm$^{-1}$. As is well known, G-band represents $E_{2g}$ mode of first order Raman Scattering in $sp^2$- C-atoms. On the other hand, the defects of graphene sheets cause D-band to appear. Another peak, i.e., 2D', observed for RGO at 2954 cm$^{-1}$, denotes second-order peak indicating combination of D- and G peaks. The 2D peak at 2687 cm$^{-1}$, implies that the nanoflakes in RGO are only a few layers thick of micron order. The characteristic modes $E_{2g}$ and $A_{1g}$ of bulk WS$_2$, arise at 354 cm$^{-1}$ and 419 cm$^{-1}$, respectively. On the contrary, in exfoliated WS$_2$ nanostructures, the concerned modes develop at 354.7 cm$^{-1}$ and 419.2 cm$^{-1}$, thus indicates the stiffening of modes. Figure 6(c) depicts Raman spectrum of WS$_2$-QDs decorated RGO cotton fabric, where modes due to WS$_2$-QDs as well as RGO are recorded. The mode frequencies and their attribution are summarized in Table 1. We notice minor shift in peak positions of

WS$_2$-QD decorating on RGO due to interfacial strain [44]. The mode analysis in case of RGO and WS$_2$-QD/ RGO are done in detail elsewhere [41].

**Table 1:** Comparison of Raman spectrum of WS$_2$-QDs obtained before and after depositing on RGO cotton fabric.

| Material | $E_{2g}$ | $A_{1g}$ | D-band | G-band | 2D-band |
|---|---|---|---|---|---|
| WS$_2$ -QD | 354.7 | 419.2 | - | - | - |
| WS$_2$-QD/RGO | 353.9 | 421.5 | 1353.1 | 1594.3 | 2989.6 |

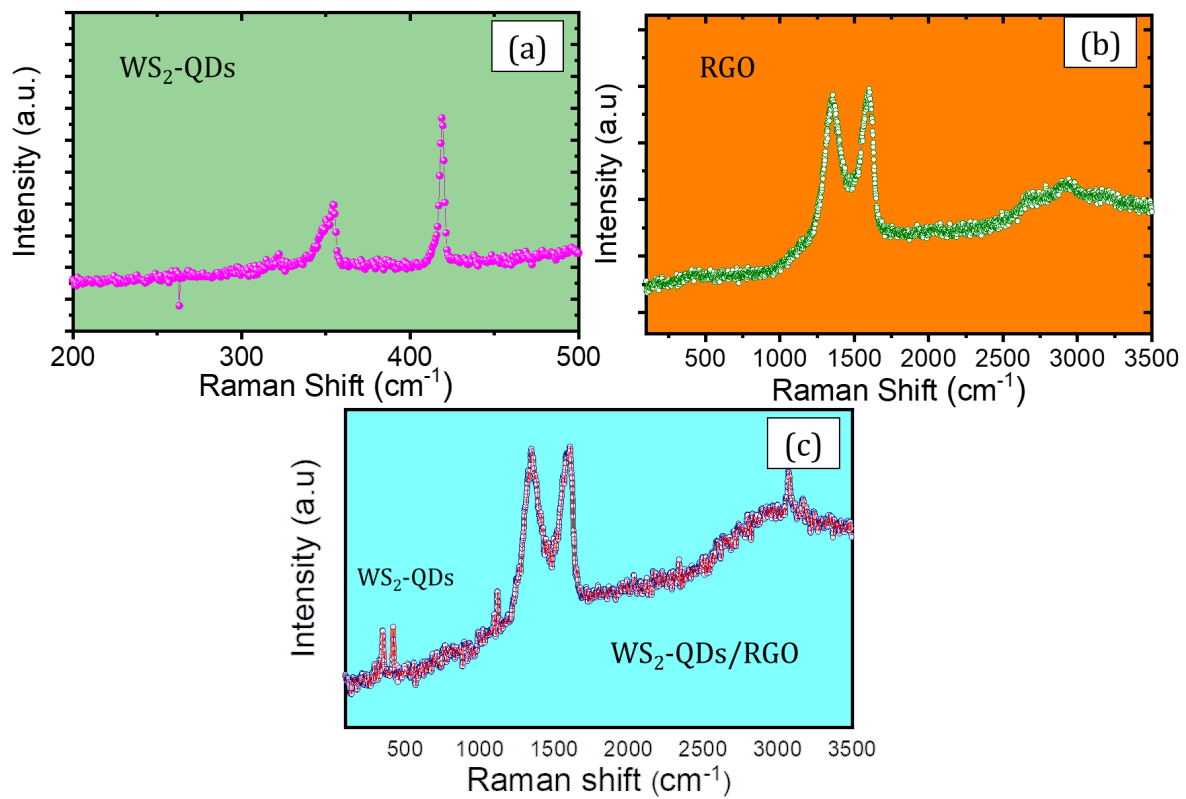

Figure 6. (a) Raman spectra of WS$_2$-QDs, (b) RGO, and (c) WS$_2$-QDs/RGO.

## 3. Heat Sensing: Role of Transport medium – Theoretical understanding

For the operation of an efficient temperature sensor, the heat transportation through any medium is a vital condition. The transport medium should possess two important physical characteristics, i.e., electronic (σ) and thermal (κ) conductivities. Thermal conductivity helps to exchange energy between lattice vibrational phonon modes and charge carriers, whereas electronic conductivity will drift the carriers to the terminal electrodes. Both κ and σ control two other important physical parameters, namely the temperature coefficient of resistance (TCR) along with thermal hysteresis ($H_{th}$). TCR is a figure of merit, such that when TCR increases, the sensor response also increases, correspondingly. TCR is associated with the mechanisms which control electronic conduction, whereas thermal hysteresis is a measure of retention of heat energy due to which the sensor fails to comeback to its initial state, therefore, it signifies the degree of sensor's efficient cycling reversibility. Both are considered to be the backbone behind the efficient performance and quantification of the sensing characteristics, e.g., sensitivity, drift, response- and recovery time.

High TCR along with low $H_{th}$ are highly desirable. Therefore, their optimum values are necessary and can be tailored through the understanding of the inter-dependency of κ and σ of the transport medium (as $WS_2$-QDs hardly occupy any space in RGO lattice), which may be accomplished through proper lattice vibrational analysis and temperature dependent electronic and thermal conductivity of the transport medium, i.e., RGO.

(a) **Phonon assisted heat transport in RGO lattice**

Thermal conductivity is an outcome of fine balance of the interaction amongst phonons generated by lattice vibration, and its scattering with different scattering agencies in the lattice. The lattice vibrational characteristics of RGO can be explained on the basis of graphene lattice and its phonon dispersion. The triangular sub-lattice structure of graphene and its phonon dispersion in the Brillouin zone [45-50] are shown in Figures 7b and 7b, respectively.

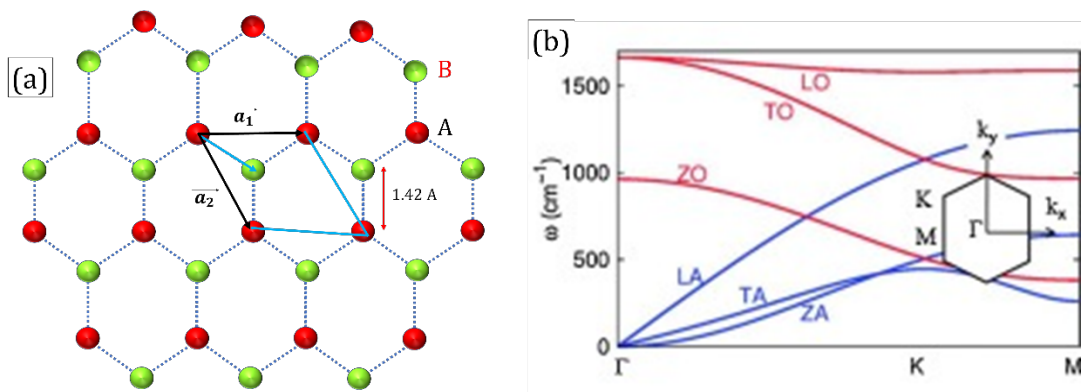

Figure 7 **(a)** Triangular sub-lattices of graphene. Each atom in one sub-lattice (*A*) has 3 close neighbors in other sub-lattice (*B*) and vice versa. **(b)** Phonon dispersion of graphene along high symmetry ΓKM lines in Brillouin zone. Inset is first Brillouin zone for the hexagonal lattice structure.

Due to two nonequivalent C-atoms in the unit cell, graphene carries six phonon branches (Figure 7b) [47, 51]. In bottom to top direction, the order of these branches is: z-directioned acoustic (ZA), transverse acoustic (TA), longitudinal acoustic (LA), z-directioned optical (ZO), transverse optical (TO), and longitudinal optical (LO) branches. A conical form appears at *K* and *K'* points of the boundary of Brillouin zone, which is like the Dirac cones of electronic structure. Nevertheless, due to in-plane modes, a non-zero phonon density of states is observed at these points [47]. Further, longitudinal (LA) and transverse (TA) acoustic modes are in-plane and propagate along basal plane with a velocity equal to that of sound waves (~20 km/s) [52]. Moreover, in pristine graphene, the LA and TA modes transfer heat from source to sink, resulting in a large thermal conductivity. Heat transport in graphene is somewhat different from that in bulk graphite's basal planes. In bulk graphite, phonons exhibit a minimum cutoff frequency ($\omega_{min}$), beyond which the heat transport approximates two-dimensional [53]. Whereas, below $\omega_{min}$, the cross-plane phonon modes exhibit strong coupling, resulting in an omnidirectional heat propagation [53]. Therefore, the combination of large phonon group velocity and large phonon mean free path (MFP) endows an ultrahigh thermal conductivity to graphene [53]. Furthermore, large phonon group velocity results from strong C-C bonds and the low weight of the C-atoms; whereas long MFP results from a special 2D phonon band structure, which makes it difficult to satisfy the phonon scattering conditions [54]. An inherently large phonon MFP, especially for the modes of longer wavelengths, yields nearly ballistic thermal transport in pristine graphene [54]. Generally, acoustic phonons, or fast traversing quanta of lattice vibration, carry heat in different materials, including graphene and in other carbon derivatives too [55]. The reason why graphene has become the promising heat flux candidate, is due to its high in-plane heat conductivity (5000 W/mK), whereas the cross-plane heat transfer is two orders smaller in magnitude [56]. Thermal conductivity '*K*' denotes a material's capability to heat conduction rapidly. Further, thermal diffusivity ($\alpha$), measures how heat propagates through the material and is generally described as:

$$\alpha = \kappa / C_p \rho_m \quad (1)$$

where $C_p$ is the specific heat, and $\rho_m$ is mass density. Instead of thermal conductivity, most of the experimental techniques measure thermal diffusivity. The picture will be somehow different when transport medium RGO is concerned. Unlike graphene, apart from variably sized graphitic regions, RGO sheets also contain residue clusters oxygenated functionalities which bond covalently to graphene surface. Thus, RGO becomes an amalgam of $sp^2$ and $sp^3$-C-atoms and nanoscopic holes [25]. Analyses also reveal that a major cause of reduced thermal conductivity of RGO is the considerable phonon scattering with oxygenated moieties which significantly deform the graphitic structure [54]. In spite of that, RGO is still considered as a 'graphene like material' because of its near graphitic properties.

Fourier's law regulating heat conduction, relates to the thermal conductivity, $\kappa$, as [48,49]:

$$\vec{q} = -\kappa \nabla T \quad (2)$$

where $q$ is the local heat flux, $\kappa$ represents thermal conductivity and $\nabla T$ denotes variation in local temperature. In RGO, phonons are the major heat carriers at high temperatures. During thermal conduction, the phonons scatter with – conduction electrons, other phonons, lattice defects, interfaces, and impurities [49,50, 53,57]. Since multiple scattering agents are present in a sample of mm length, the phonon MFP ($l$) becomes small; as a result, the heat conduction follows Fourier's law and is mainly diffusive. For $l > L$, phonons do not encounter scattering while travelling through the sample and the heat transport becomes ballistic [48-50]. Thus, RGO possesses a greatly reduced thermal conductivity than graphene [47-50].

In this context, after taking into consideration of steady state heat diffusion, Eq. (2) becomes:

$$\kappa \cdot \nabla^2 T(\vec{r}) + q(\vec{r}) = 0 \quad (3)$$

The mechanism of heat transfer in a $WS_2$-QDs/RGO heterostructure is via the coupling of electrons (charge carriers) in $WS_2$-QDs and RGO and transport along the RGO maize. This is, therefore, the predominant mechanism of heat transfer in $WS_2$-QDs coated RGO.

**(b) Temperature dependent electrical conductivity (σ) of heat Transport medium**

Unlike graphene, RGO sheets contain large number of flakes having mono/few-layered graphene sheets, with a thickness from micron to nm. Each sheet contains a mixture of $sp^2$- and $sp^3$- graphitic regions and clusters of residual oxygenated functionalities covalently bonded to the graphene surface [25]. Theoretical investigations suggest that complete reduction of GO to RGO and then to pure graphene, is difficult to achieve; and lattice damage is inevitable [58]. As a result, 20-30 % or even more (depending on the synthesis route) residual functional elements present on the basal plane and edges [25]. As of now, the interpretation of electrical properties of RGO is based on its disordered entity [59, 60].

In a disordered semiconductor, electric transport is occurring mainly via extended states in conduction band, valence band and mobility gap [61]. If there are localized states near Fermi energy, the probability '$P$' of carriers 'jumping' from one localized state to another with higher energy, is determined from three factors [62]: (a) attempt frequency ν (ph), between $10^{12}$-$10^{13}$ s$^{-1}$, (b) electron (hole) wavefunction, and (c) probability of finding a phonon with excitation energy ($w$) large enough to hop [62]. The hopping conductivity is found to occur either from Nearest Neighbor Hopping (NNH) or Variable Range Hopping (VRH); with the assistance of a phonon in both cases. NNH is the most probable form of hopping and a thermal activation-based temperature dependent conductivity is observed [59, 62-66]. Depending on its energy, a carrier can move from one site to another, such that these sites lie at different distances. The resultant conductivity deviated from NNH to VRH [59, 62-66]. The hopping distance R increases with decreasing temperature. Thus, NNH dominates at elevated temperatures, while at lower temperatures, VRH is usually expected. Therefore, the carrier transport phenomenon is to be understood from the temperature dependent electronic conductivity, as this dependence indicates the underlying transport

mechanism. In wide temperature ranges, in disordered materials, the direct current (DC) conductivity is [59]:

$$\sigma = \sigma_0 e^{-\left(\frac{T_0}{T}\right)^p} \quad (4)$$

where $\sigma_0$, a pre-exponential factor, is decided by the underlying system; $p$ is power exponent and is material specific and may also have dependence on operating temperature range; and $T_0$ is characteristic temperature. Generally, a hopping conduction model perfectly describes the temperature dependence of conductivity. There exists a critical temperature, above which Mott variable range hopping (M-VRH) based conduction, with $p = 1/4$, dominates the carrier transport, while the carrier hopping is determined by the amount of energy [59, 62-66]. Further, below critical temperatures, a Coulombic interaction is usually observable among the charge carriers, as explained by Efros-Shklosvkii variable range hopping (ES-VRH), with $p = 1/2$ [59, 62-66]. Schematic of charge transport in RGO having flakes of the order of sub-micron thickness, showing hopping (intra-flakes) and tunneling transport (inter-flakes) mechanisms is given in Figure 8(b-c).

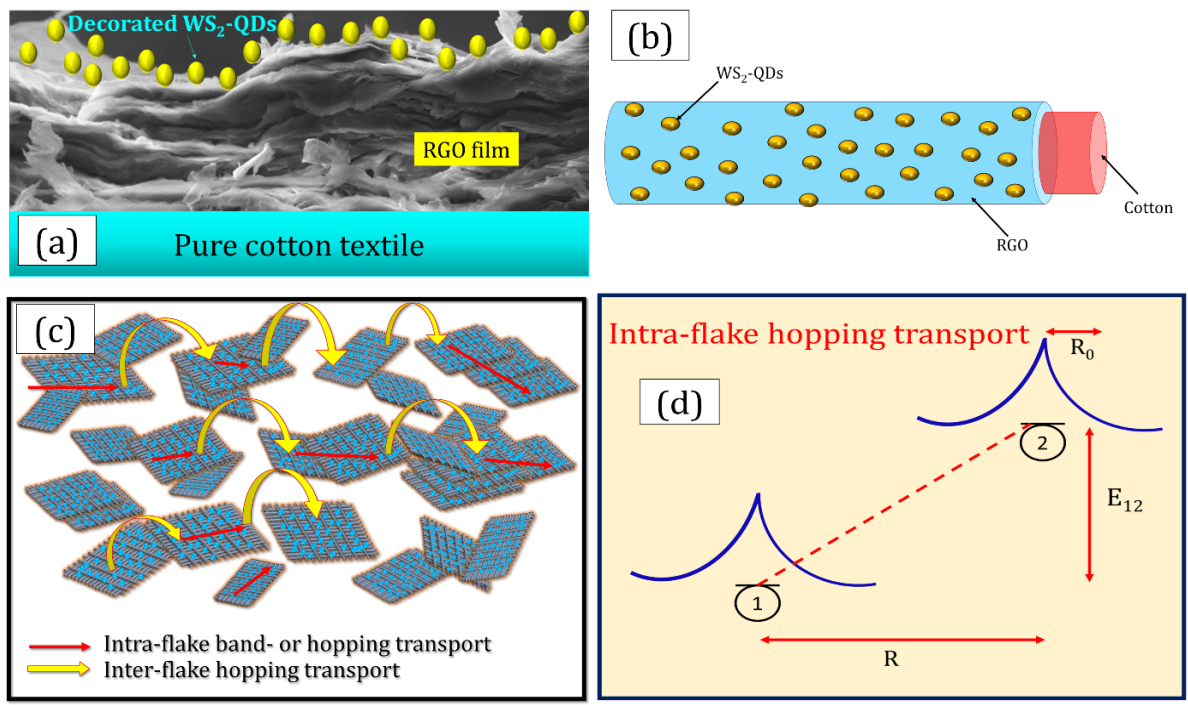

**Figure 8**. (a) Stacking of RGO sheets and decoration of WS$_2$-QDs on cotton cloth, (b) Schematic of the fabrication of WS$_2$- QDs decorated RGO film, wrapped on a cotton cloth, (c) RGO flakes are in random orientation where inter-flakes transport by carrier tunnelling,

and intra-flakes carrier transport by VRH is shown, and (d) in case of intra-flake, hopping is shown between the two localized states separated by distance R and energy $E_{12}$; and $R_0$ is the localization length.

The morphology of RGO film is indicated in Figure 8(a-b), where a random orientation of 2D flakes within the plane, and $WS_2$-QDs decoration on RGO flakes are clearly perceived. Depending upon the intrinsic electronic structure and delocalization of defect states, charge transport within an individual flake can be either diffusive (band) or hopping-type. Primarily, a variable range hopping is expected with hopping probability varying as $\sim\exp[(-2R/\alpha)-(W/k_BT)]$, where $R$, $\alpha$, and $W$ represent the distance between nearest neighboring flakes, attenuation length, and energy difference between conduction states within the flakes, respectively, and $k_B$ is the Boltzmann constant (Figure 8c) [59].

## 4. Temperature Sensor Fabrication and Testing setup

Temperature sensor of size 10 mm × 10 mm was fabricated from the $WS_2$-QDs/RGO heterostructure. $WS_2$-QDs were decorated on the top surface of the RGO coated textile. Electrodes were fabricated at the two ends of the film with Ag paste to act as terminal electrodes. Temperature sensing experiments were carried out using a two-probe system (Linkam, model T-95 PE) fitted with a temperature controller (Figure 9) to set the targeted temperature. The heating/cooling derived sensing response was obtained in 77-398K temperature range where data acquisition was made on a Semiconductor Characterization System (Keithley 4200 SCS).

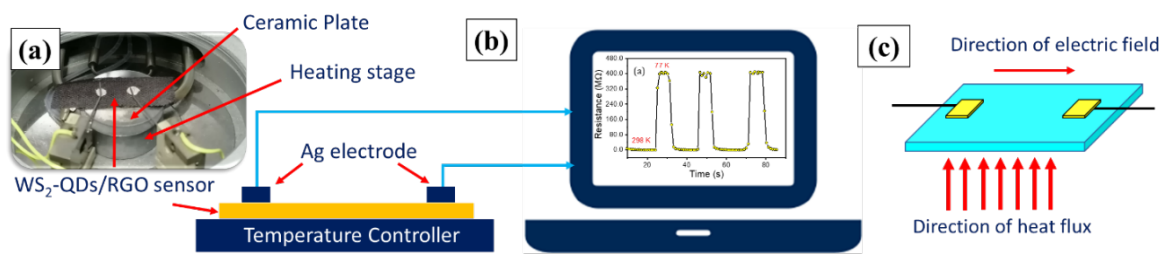

Figure 9. (a) Digital image of $WS_2$-QDs/RGO e-textile film, (b) scheme of experimental conditions to monitor temperature dependent sensor response, and (c) directions of heat flux and electric field.

To measure temperature induced resistance, change in $WS_2$-QDs/RGO film, a dc bias of 12 V was applied to collect free excess carriers at terminal electrodes of the sensor. Figure 9(c) illustrates the directions of heat flux and electric field, which are usually located in cross plane geometry during sensing measurement. On heating, $WS_2$-QD injects electrons (e⁻) as excess carriers in the disordered RGO lattice [41] wherein multiple events take place such as – carrier scattering with multiple types of scattering agents [25, 41, 67-68], trapping and de-trapping with defect centres, recombination losses with e-h recombination during its journey, etc., and finally those carriers able to drift under electric field to make their shortest journey, will reach to the terminal electrodes after certain transit time, $t_r$. These are schematically demonstrated in Figure 10. The cross section of the scattering processes further determines the carrier transit time ($t_{tr}$). Furthermore, both $t_{tr}$ and heat diffusion process together decide how fast the device reads a particular temperature; which then, further depends upon these very scattering phenomena. Thus, diffusion and drift processes occurring concurrently in cross plane geometry influence the working of a temperature sensor or a thermometer. Phonon mean free path ($\lambda_{ph}$) and $t_{tr}$ are mutually dependent on each other. Thus, for efficient operation of a temperature sensor, a trade-off occurs in between by controlling the scattering factors as well as the applied dc bias.

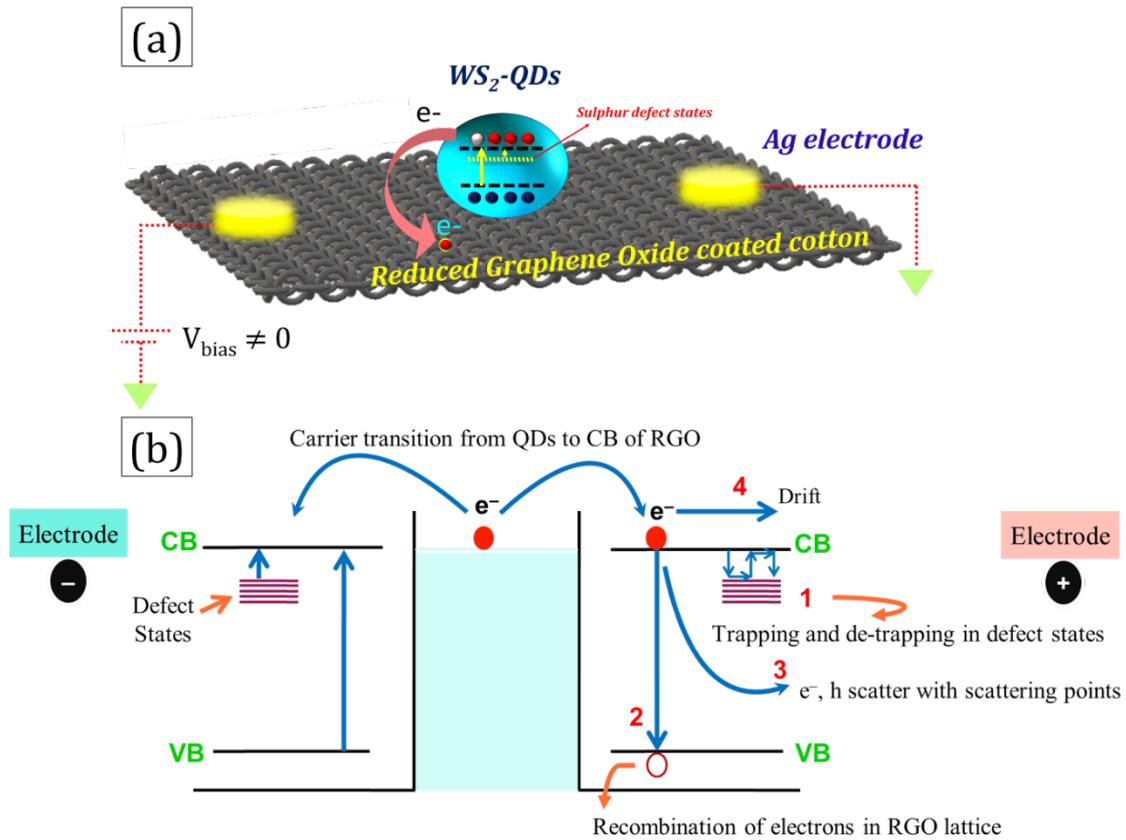

Figure 10. Schematic diagram depicting excess carrier injection from WS$_2$-QDs, carrier scattering, carrier loss, etc., while transportation in the transport medium (RGO) to reach to the terminal electrodes.

## 4.1 Experimental Results

Sensing experiments were conducted in two modes – one i.e. in conventional static mode where the sensor was placed on the sample chuck, and it was slowly heated to attain desired/ targeted temperature by selecting increment/decrement ramp which decides the rate of increase in temperature of the chuck depends on the temperature. In case of instant mode, we first achieved the desired temperatures in an insulated chamber by various means such as a liquid nitrogen/heating coil. Once the desired temperature is reached, the sensor directly put inside the targeted temperature chamber in instant manner. The sensor responds to the sudden change in temperature from room temperature to the targeted temperature.

### 4.1.1 Sensors' Parametric Analysis in conventional static mode

All experiments were performed under ambient conditions, until stated otherwise. We investigated the temperature sensing response by monitoring the change in resistance over a broad temperature range (77 K to 398 K). Typical I-V characteristics, shown in Figure 11(a), provides an excellent ohmic interface between the WS$_2$-QDs/RGO film and the Ag electrodes. Further, this ohmic nature is found temperature independent. The slope analysis of I-V plots shows abrupt increase in resistance of the sensing element at low temperatures (below 298 K); where the resistance becomes of the order of MΩ-GΩ. The temperature dependent resistance plot (Figure 11(c)) of the sensing element bare RGO and WS$_2$-QD/RGO, shows both the elements are semiconductors. During cycling test (Figure 11(d-e)) of both the sensors at various temperature ranges, the temperature was first raised to a target value and kept for 2 min to ensure thermal equilibrium condition between sensor and sample holder chuck, followed by recording of typical response, and then the sensor was allowed to cool down to bring back to its initial state. One such repeated cycling test in the temperature range 298-323 K (25 -50 °C) is demonstrated in Figure 11(f). While cycling in that temperature range, the change in resistance of bare RGO coated cotton sensor and WS$_2$-QD/RGO sensor are ~ 20% and ~ 30 % respectively (Figure 11(d-e)) i.e. the response is nearly 50% higher in the latter case; and the difference in response is found temperature range dependent.

While doing parametric analysis, temperature coefficient of resistance (TCR) comes first as it decides the quantitative values in most of other parameters concerned to a temperature sensor. A series of static cycling test was conducted in various high temperature as well as low temperature ranges (Figure 11(g-h)) and the TCR and thermal hysteresis (Hth) values were calculated from the expression(s):

$$TCR(\%) = \frac{1}{R_0} \frac{\Delta R}{\Delta T} \cdot 100\% \qquad (5)$$

$$H_{th} = \frac{R_0 - R_f}{R_0} * 100 \qquad (6)$$

Where, $R_0$ = initial/reference resistance of device; $\Delta R$ = change in sensor resistance with $R_0$ being the reference resistance, and $\Delta T$ = corresponding temperature change. TCR was directly calculated by monitoring sensor response induced on shifting the sensor from room temperature to the targeted temperature. It is found that WS$_2$-QD/RGO sensor showed excellent TCR vis-à-vis bare RGO in all most all temperature ranges where superior values are noticeable in the low temperature ranges (Figure 11(g-h)); most likely due to the hopping transport mechanism in the sensing elements as described earlier. On the other hand, thermal hysteresis loss is insignificant in the low temperature ranges and negligibly small upto 5.2 % upto 398 K (Figure 11(i)). As expected, the presence of high TCR and low $H_{th}$ directly reflected the superiority of WS$_2$-QD/RGO sensor with sensitivity, response time, recovery time, and resolution measured in static mode and are summarized in Table II.

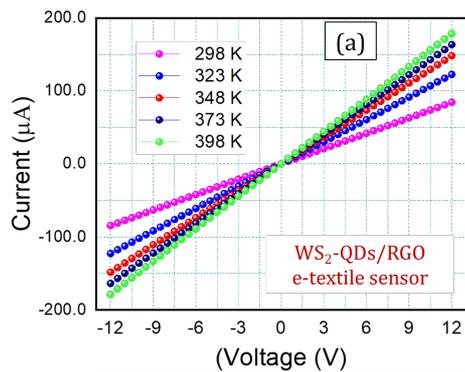
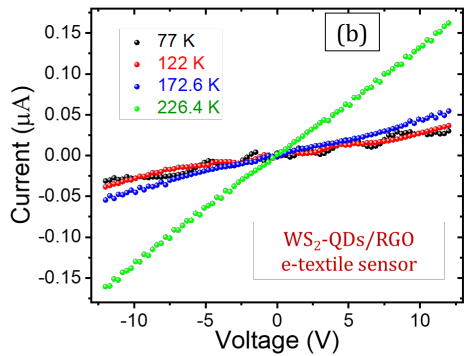
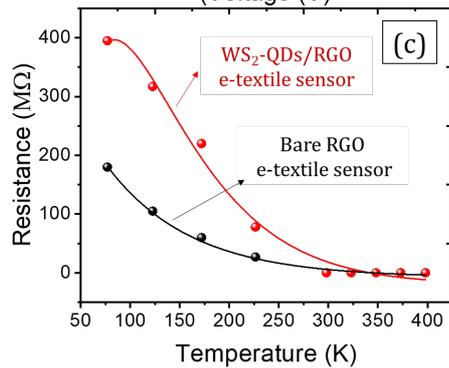
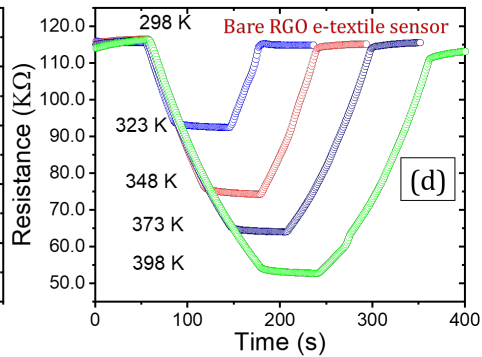
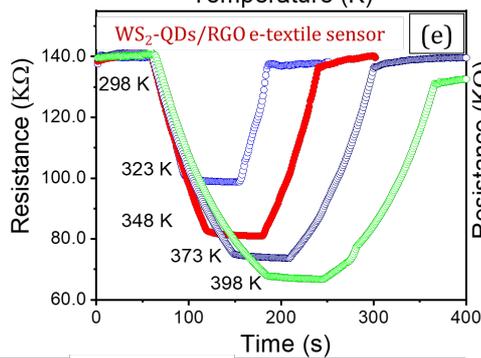
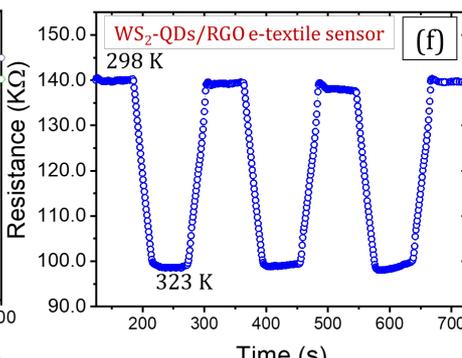
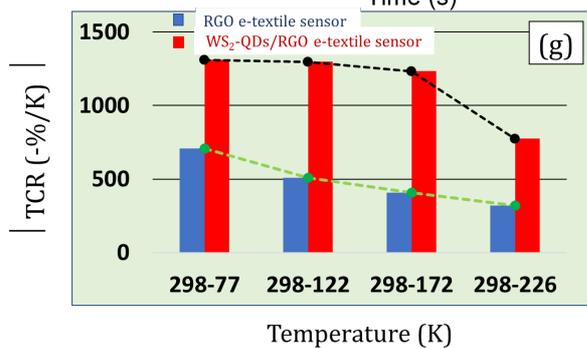
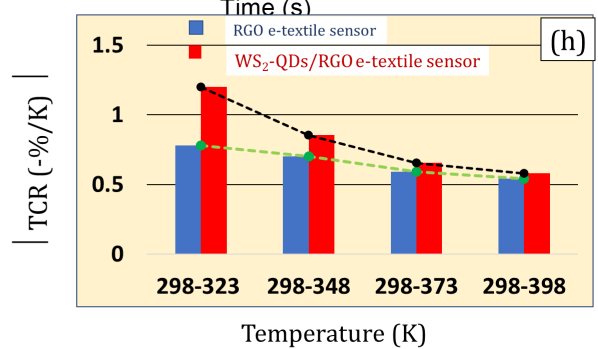
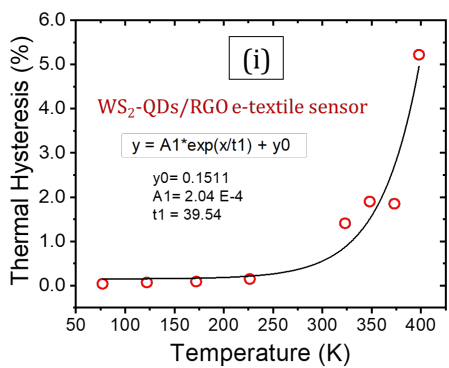

Figure 11. (a-b) I-V characteristics at high and low temperature for WS$_2$-QD/RGO sensor, (c) Resistance versus temperature plot in the temperature range 77 K -398 K for bare RGO and WS$_2$-QD/RGO sensor, (d-e) Static resistance vs time plot for various temperature range for WS$_2$-QD/RGO and bare RGO sensor, (f) Cyclic resistance vs time response in 298-323 K temperature range in case of WS$_2$-QD/RGO sensor, (g-h) TCR bar graph of both the sensors for comparative analysis at various temperature ranges from 77-398 K, and (i) Thermal hysteresis plot as a function of temperature for WS$_2$-QD/RGO sensor.

### 4.1.2 Resolution Test in static cycling measurement

Additionally, dynamic response of WS$_2$-QD/RGO sensor was also examined as a trial example in 298-373 K temperature range (Figure 12). From Figure 12(b), the average change in resistance between 298-373 K is ~925 Ω for a 10 K step change in temperature. If the minimum measurable signal strength is considered to be ~100 mΩ, the resolution turns out to be 0.01 K, i.e., 0.01 K change in temperature can be measured. This ascertains ultrahigh sensitivity of device. Besides, the change in resistance versus the change in temperature reveals approximately linear relationship. Therefore, standardization and calibration may not be an issue.

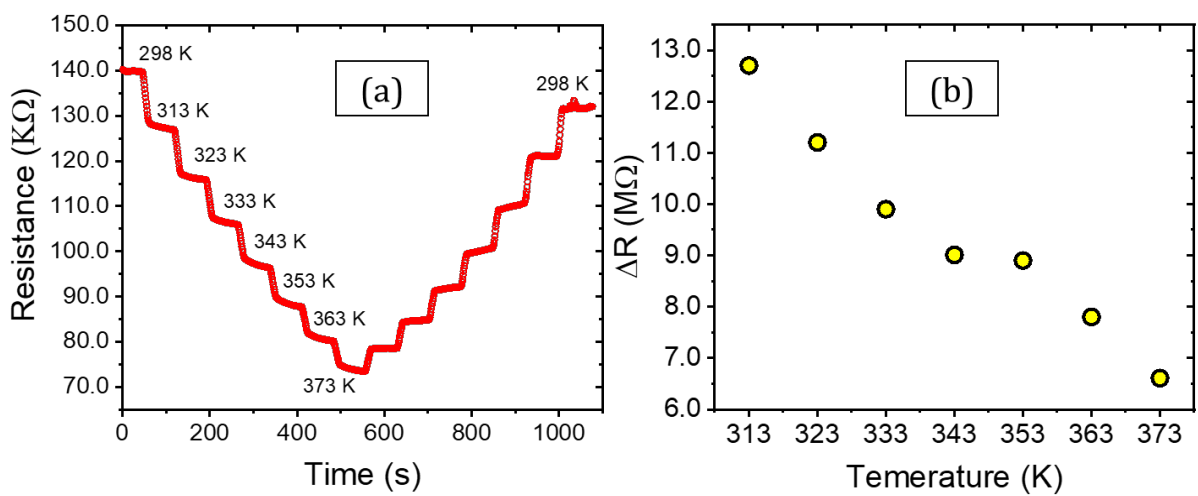

Figure 12. (a) Resolution test were done in conventional static mode. The variation in resistance with temperature measured frm dynamic plot, (b) sensor response in 298-373 K for a ~10 K step change in temperature.

**Table II: -** Sensing parameters measured at high and low temperatures with respect to room temperature as a reference.

| Temperature Range | Initial Temperature (K) | Final Temperature (K) | TCR (%/K) | Thermal Hysteresis (%) | Response Time (s) | Recovery Time (s) |
|---|---|---|---|---|---|---|
| From RT to high temperature | 298.0 | 398.0 | -0.58 | 5.22 | 92.65 | 144.13 |
| | 298.0 | 373.0 | -0.65 | 1.85 | 70.81 | 71.39 |
| | 298.0 | 348.0 | -0.85 | 1.90 | 49.36 | 50.04 |
| | 298.0 | 323.0 | -1.20 | 1.41 | 27.17 | 26.34 |
| From RT to low temperature | 298.0 | 226.4 | 775 | 0.15 | 67.30 | 68.40 |
| | 298.0 | 172.6 | 1233 | 0.09 | 106.2 | 111.00 |
| | 298.0 | 122.0 | 1298 | 0.07 | 150.00 | 137.8 |
| | 298.0 | 77.0 | 1310 | 0.05 | 202.20 | 155.34 |

**4.2 Real-time performance - instant sensing measurement like thermometer**

Instantaneous temperature measurements like thermometer were conducted in the low temperature as well as high temperature sides vis-à-vis the ambient temperature (Figure 13), where sensor was moved swiftly from reference to targeted temperature in an instant manner. The unexpected increase in TCR value demonstrate the superiority of the $WS_2$-QD/RGO sensor in terms of sensitivity and resolution. Within each range, the rate in resistance change occurs in a linear fashion; which has further advantage where device does not require further rigorous calibration exercise. Moreover, there was no apparent shift in sensor baseline over cycling (Figure 13(b)), indicating, heat retention loss is virtually nil and thus, thermal hysteresis is not an issue. The TCR and $H_{th}$ values in low temperature ranges are tabulated in Table-III along with response- and recovery-time.

In addition, instant sensing tests were also carried out at high temperature ranges in between 298–398K in a home-built insulation chamber (Figure 13(e)) fitted with variable heater circuitry fitted with digital read-out electronics, and a window for inserting sensor in an instant manner inside a readymade targeted stable thermal environment. The instant test results about TCR, $H_{th}$, response- and recovery time for the entire temperature range 77-298 K are impressive and incorporated in Table III. Response- and recovery times are found temperature range dependent; very fast at the low temperature ranges vis-à-vis the higher temperature side. It is obvious that the scattering events as discussed in section 3.0 have taken place; and becomes more with the increase in temperature leading to higher carrier transit time. In brief, the resultant outcome is long response- and recovery time.

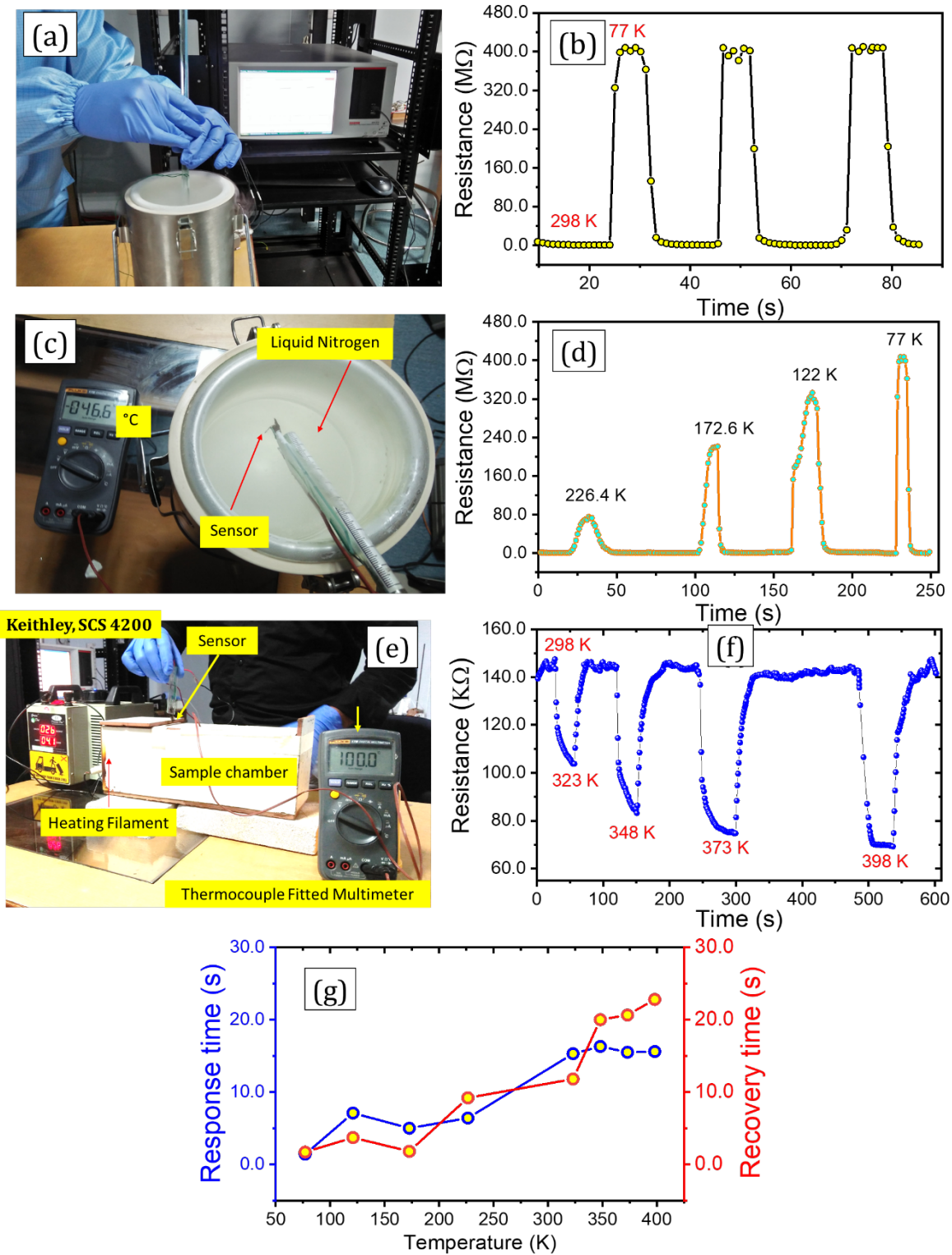

Figure 13. (a) Digital photograph of the experimental set-up, (b) Cyclic response of WS$_2$-QDs/RGO sensor in 298 to 77 K temperature range. This is performed in repetition once by dipping the sensor in liquid N$_2$ and subsequently removed it from the nitrogen bath, (c) Digital photograph of sensor operation at different target temperature and the process is same

as mentioned in case of (a), (d) Resistance vs time as a function of various low temperature ranges.  (e) Digital photograph of experimental set-up in convection mode (f) Sensor response to instant variation of different temperature ranges, and (e) response- and recovery time as different temperature ranges.

**Table III: Instant temperature measurement - Sensing parameters measured at high and low temperatures with respect to room temperature as a reference.**

| Temperature Range | Initial Temperature (K) | Final Temperature (K) | TCR (%/K) | Thermal Hysteresis (%) | Response Time (s) | Recovery Time (s) |
|---|---|---|---|---|---|---|
| From RT to high temperature | 298.0 | 398.0 | -0.56 | 0.01 | 15.6 | 22.8 |
|  | 298.0 | 373.0 | -0.64 | 0.70 | 15.5 | 20.65 |
|  | 298.0 | 348.0 | -0.85 | 0.01 | 16.3 | 20.0 |
|  | 298.0 | 323.0 | -1.16 | 0.10 | 15.3 | 11.8 |
| From RT to low temperature | 298.0 | 226.4 | 713 | 0.02 | 6.4 | 9.2 |
|  | 298.0 | 172.6 | 1248 | 0.01 | 5.0 | 1.9 |
|  | 298.0 | 122.0 | 1295 | 0.03 | 9.3 | 3.7 |
|  | 298.0 | 77.0 | 1315 | 0.05 | 1.4 | 1.7 |

**4.2.1 Human Trial for monitoring body temperature**

Human trial is conducted to further determine the sensor's ability to estimate human body temperature. For this purpose, the sensor was placed directly on the hand for certain time (~ 4-5 sec) and then detached from the hand. This process was repeated many times to ascertain the repeatability of sensor response. Figure 14 reveals that the response is quite stable and repeatable too. The body temperature was assumed to be ~310 K vis-à-vis the room temperature i.e., 298K, where the change in resistance was found to be ~18 k$\Omega$. For proper understanding of the graph, temperatures are also mentioned in the figure. A simple resolution calculation is done considering minimum measurable signal strength of 100 $\Omega$ (which is very much at higher side; ordinary sensors measure milli- to micro scale change). In this case, then the resolution value stands at 0.06 K, implying the minimum change in temperature that can be measured with this sensing device. Sensing characteristics for this trial test are given in Figure 14 (c).

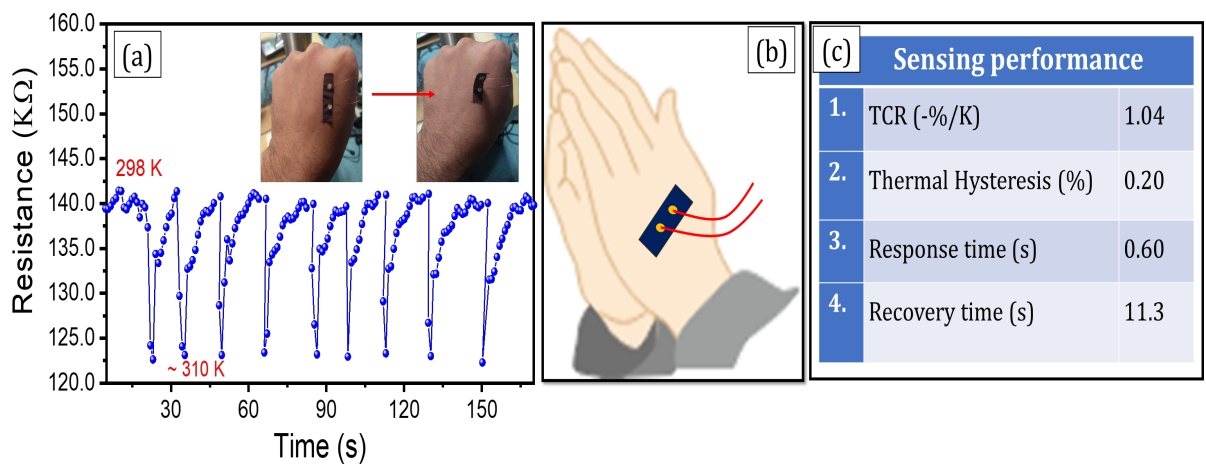

Figure 14. (a) Instantaneous body temperature measurement and its temperature cycling response shown by the developed sensor. Inset picture shows the physical demonstration), (b) Schematic representation of sensor placed on the hand, and (c) sensing data table.

## 5.0 Device flexibility tests

To substantiate our claim for flexibility as well as wearability of developed sensor, mechanical deformation tests such as bending, twisting as well as stretching were conducted to evaluate its ability to absorb mechanical deformations without any considerable impact on temperature sensing performance of $WS_2$-QD/RGO heterostructure.

**5.1 Bending Test**: The impact of bending on device characteristics was evaluated by monitoring the sensor baseline resistance under different bending conditions. Insets from I-IV in Figure 15A show digital images of the sensor bent at various angles (θ): ~$0^0$, $15^0$, $30^0$, and $85^0$, respectively. To ensure a constant thermal flux crossing the sensor under all bending angles, the sensor was attached to a curved hot surface (Figure 15 A(b-c)), and the sensor response was monitored at a fixed dc bias. The sensor exhibited great flexibility by easily taking the shape of the curved tub. The results revealed no appreciable change in sensor baseline upon bending.

**5.2 Twisting Test:** For twisting related deformities, both the ends of the sensor were clamped and the sensor response (in terms of baseline resistance) was measured in three different positions: without twist, half twist and full twist. The tests results included in Figure 15B reveal no apparent change in sensor baseline resistance value, indicating the flexible nature of the developed sensor.

**5.3 Stretching Test**: For stretching tests, one end of the developed sensor was fixed, and manually pulling the other end. During the stretching, simultaneous *in-situ* resistance continuity check was performed across the sensor terminals. Since, our e-textile is a cotton cloth, which does not exhibit stretching properties like polymers, a maximum stretching of ~1 mm could be achieved at hard pulling. The experimental studies like resistance continuity check before and after stretching test as shown in Figure 15C, indicate no deterioration in sensor performance.

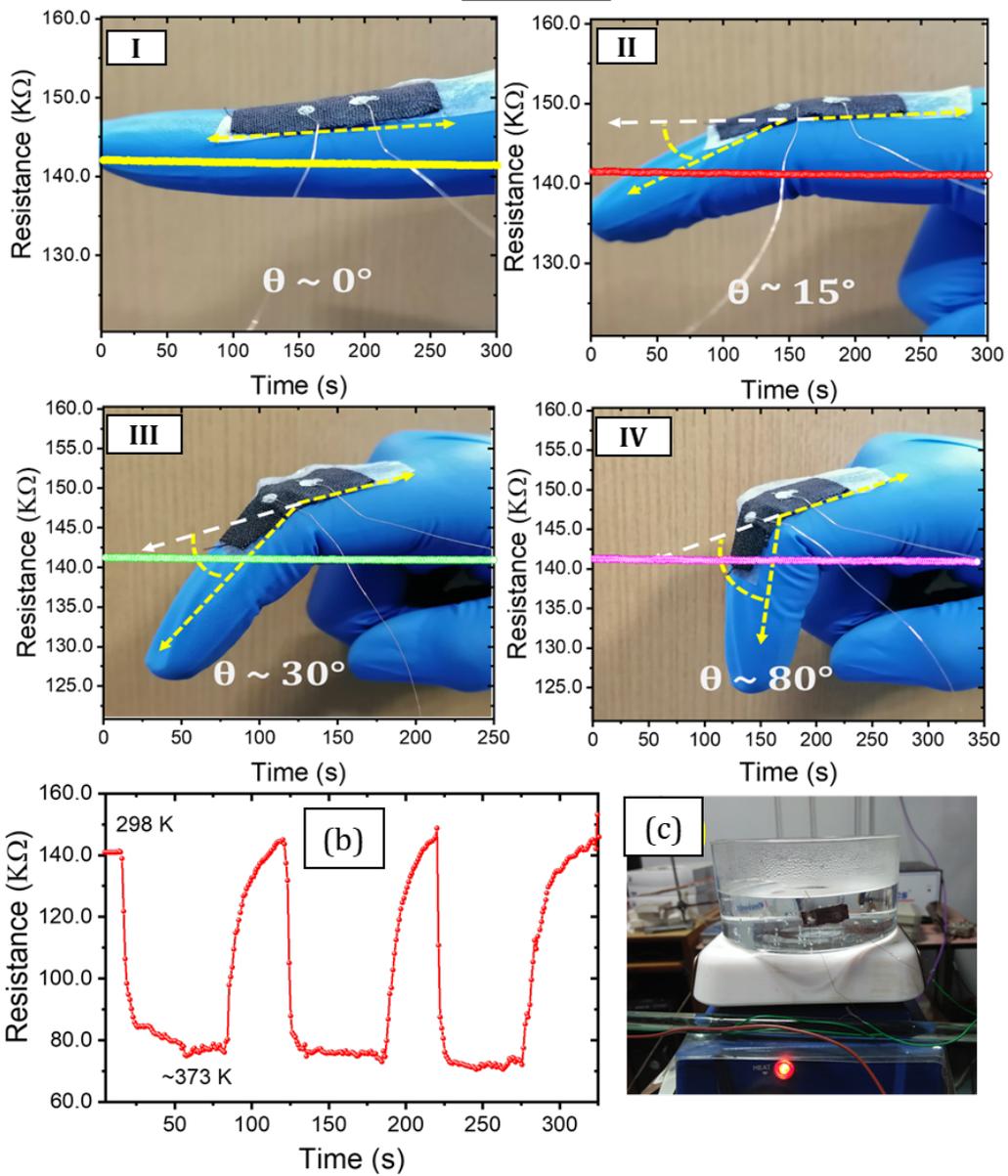

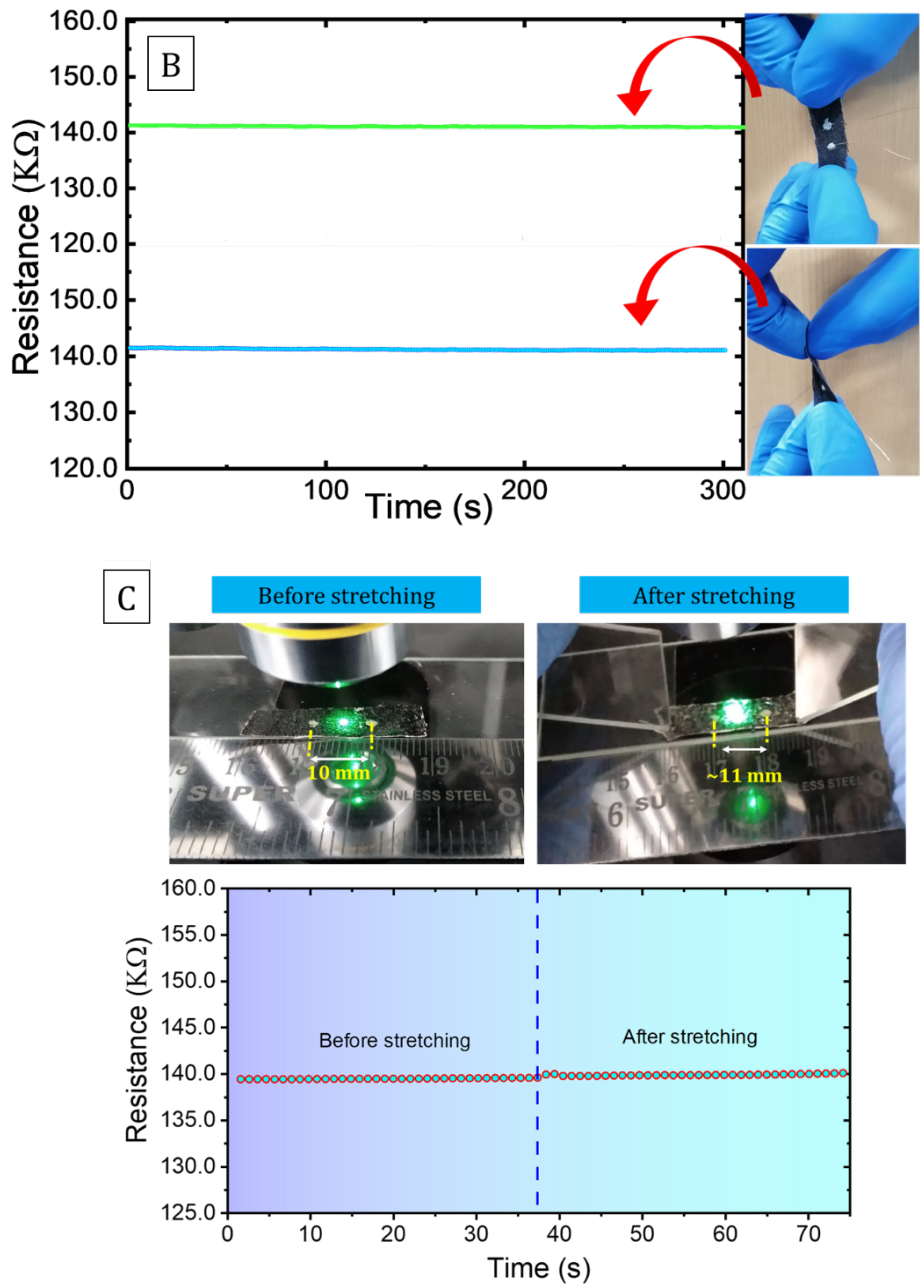

Figure 15 **A. (a)** Baseline resistance versus time of WS$_2$-QDs/RGO sensor for different bending angles, **(b-c)** cyclic temperature response of sensor by fixing it to a curved hot surface of the quartz tube as visible in the photograph (c). **(B)** baseline resistance fluctuation check during twisting, **(C)** base level resistance fluctuation check during stretching.

A comparative analysis of the developed sensor and the temperature sensors reported so far is presented in Table IV for the interest of the researchers.

Table IV: Comparison of temperature sensing performance of the fabricated WS$_2$-QDs/ RGO sensor with some of the reported temperature sensors.

| Materials | Temperature range (K) | TCR (%/K) | Other parameters | Source |
|---|---|---|---|---|
| GNWs@PDMS substrate | 308-318 | 21.40 | Response time = 1.6 s<br>Recovery time = 8.52 s<br>Resolution = 0.1 K | [33] |
| RGO nanosheets/elastomeric PU composite | 303-353 | $\Delta R/\Delta T$ = 1.34% | Resolution = 0.2 K<br>Cyclability = 10000 cycles | [68] |
| SWCNT TFTs with PANi nanofibers | 288-318 | -1 | Cyclability = 1000 cycles | [69] |
| CNT/PEIE/PDMS composite@PET substrate | 298-313 | ~0.85 | — | [70] |
| PEDOT:PSS@paper substrate based thermistor | 263-298 | -3 | — | [71] |
| MWCNTs/alumina composite self-standing films | 303-373 | -0.96 | Response time = 40 s<br>Recovery time = 185 s<br>Thermal hysteresis = 0.62% | [72] |
| V$_2$O$_5$ thin films | 319-343 | ~-4 | — | [73] |
| RGO/alumina composite self-standing films | 300-77 | -1999.8 | Response time = 0.3 s<br>Recovery time = 0.8 s | [25] |
| | 296-373 | -0.98 | Response time = 3.96 s<br>Recovery time = 6.01 s | |
| WS$_2$-QDs/RGO coated e-textile (Instant Measurement) | 298-77 | 1315 | Response time = 1.4 s<br>Recovery time = 1.7 s | Present work |
| | 298-398 | -0.56 | Response time = 15.6 s<br>Recovery time = 22.8 s | |

# Conclusion

A flexible and wearable temperature sensor is fabricated by decorating $WS_2$-QDs on RGO-coated e-textile fabric ($WS_2$-QDs/RGO). The fabrication steps presented here is facile, scalable, and economic. In the fabricated sensor, $WS_2$ injects excess electrons into RGO medium to be transported with high mobility and fast transit time. Temperature sensing performance of the developed device were investigated to assess the effectivity as well as utility of $WS_2$-QDs and RGO heterostructure. The measured sensing data in the high temperature range (298-398K) are: TCR ~-0.56 %/K, hysteresis $H_{Th}$ ~ 0.01%, response-recovery time ~15.6 s, and ~22.8 s, and resolution ~0.01K; whereas improved data were obtained for sensing performed in low temperature range (298 - 77K). The data recorded in both cases are comparable and sometimes superior to other reported sensors for conventional static measurements. When the sensor was employed for instant temperature measurement just like a traditional thermometer, faster response with enormously improved performance sensor parameter were achieved, as listed in Table-II and III. Human trial was conducted to measure body temperature similar to the thermometer and the sensor was found to be able to measure a minimum change in temperature of 0.06K, though the minimum measurable signal strength is taken at exorbitantly higher side vis-à-vis the conventional solid-state sensors. The results reveal the efficacy of developed sensor in terms of high sensitivity as well as resolution. Temperature sensing mechanism is explained in terms of two important material characteristics, i.e., thermal and electrical conductivity of the transport medium; since their inter-dependency controls TCR, and $H_{th}$. Both these parameters fine-tune all other sensing parameters. The impact of mechanical deformation was studied on bent, twisted, and stretched samples. Results revealed no considerable change(s) in sensing parameters, thereby complimenting our claim of flexibility of the fabricated device. Further research by our group is being conducted to study the effect of QDs size, its density as well as homogeneity in distribution in RGO medium, on the performance of $WS_2$-QDs/RGO based temperature sensor.

**\*Note:** The characterization results discussed in Section 2, have already been reported for $WS_2$-QDs/ RGO heterostructure based photodetector, in our paper entitled "$WS_2$ Quantum Dots on e−Textile as a

Wearable UV Photodetector: How Well Reduced Graphene Oxide Can Serve as a Carrier Transport Medium?", published in ACS Applied Materials & Interfaces (doi: 10.1021/acsami.0c08028). For the convenience of readers and to maintain continuity of the present paper, we have reproduced the characterization results here also, with permission from the publisher.